\documentclass[prl,aps,twocolumn,showpacs]{revtex4}
\usepackage[dvips]{graphicx}
\usepackage{dcolumn}%
\usepackage{amsmath}%
\usepackage{amsfonts}%
\usepackage{amssymb}
 \usepackage[usenames]{color}
\usepackage{appendix}
\providecommand{\U}[1]{\protect\rule{.1in}{.1in}}
\newcommand{\be}{\begin{equation}}
\newcommand{\ee}{\end{equation}}
\newcommand{\bea}{\begin{eqnarray}}
\newcommand{\eea}{\end{eqnarray}}
\newcommand{\bt} {\begin{tabular}}
\newcommand{\et} {\end{tabular}}

\newcommand{\ds}{\displaystyle}
\newcommand{\ba} {\begin{array}}
\newcommand{\ea} {\end{array}}
\begin{document}

\title{The effect of light scattering in cavity electrodynamics: Fresnel equations with decoherence}

\author{  Natalya A. Zimbovskaya$^1${\footnote{Corresponding author: nzimbovskaya@gmail.com}} and Abraham Nitzan$^{2,3}$}

\affiliation
{Department of Physics and Electronics, University of Puerto Rico-Humacao, CUH Station, Humacao, PR 00791, USA}  
\affiliation{$^2$Department of Chemistry, University of Pennsylvania, Philadelphia, PA19104, USA}
\affiliation{$^3$School of Chemistry, Tel Aviv University, Tel Aviv, Israel}

\begin{abstract}

We consider the effect of  light decoherence in a Fabri-Perot (FP) microcavity. We show that, when present, it may significantly change its linear response  to the incident light.The optical properties of the bounding mirrors are described using the classical electrodynamics, and the light scattering causing the phase breaking is treated as a multichannel scattering problem employing a model introduced by Büttiker to describe dephasing in electron transport. We show that the dephasing causes a gradual erosion of the cavity photon mode and thus impedes the formation of molecular polaritons. 
 Consequently, the polaritons signatures in the transmission and absorption optical spectra of the microcavity are fading away as the scattering intensifies.  
\end{abstract}

\date{\today}

\maketitle
\date{\today}
\maketitle

\subsection{I.Introduction}

The collective response of a molecular ensemble to incident light is a repeatedly studied theme in the research of light-matter interaction with manifestations ranging from molecular excitons \cite{1,2}, optical properties of molecular aggregates \cite{3,4,5,6}, exciton-plasmon and exciton-cavity modes interactions at molecule-metal interfaces \cite{7,8,9} and the emergence of hybrid light-matter modes (polaritons) under strong molecules-radiation field coupling. Common setups for investigating some of these phenomena are Fabri-Perot (FP) cavities where molecules are placed in between two mirrors or plasmonic cavities where they occupy gaps between metal particles \cite{10}. Observations and theoretical treatments show significant effect of the cavity environment on rates of chemical reactions \cite{11,12,13,14}, energy transfer intensities \cite{15,16,17,18,19} as well as novel modes of behavior such as polariton Bose-Einstein condensation \cite{20,21,22,23}. 

Importantly, the strong coupling behavior often expresses the collective coupling of many molecules and not the (much weaker) coupling of a single molecule to the radiation field. Recognizing this, it is natural to question the effect of static and dynamic disorder on this collective response \cite{24,25,26}. In particular, since thermal motion always exists in (mostly room temperature) experimental setups, understanding the interplay between collective dynamics and dephasing is needed for understanding the optical response of these systems. Several recent theoretical treatments are mostly based on extending the Tavis-Cummings (TC) type models of molecules interacting with one or more cavity modes by making  one of the model parameters  describing the molecular subsystem a stochastic function of time \cite{27,28,29,30}. Alternatively, for electronic strong coupling the molecular nuclear dynamics may be included, leading to Holstein-Tavis-Cumming model for a more detailed cavity vibrionic dynamics \cite{31}

While the above models have been very useful in revealing important aspects of cavity electrodynamics, they can not account for all aspects of cavity response because they include the cavity only implicitly via its modes. To emphasize this point consider the linear response to light incident from the far field interacting with the cavity mode(s) which in turn are coupled to the molecular subsystem and this interaction modifies the standard optical observables - transmission, reflection and absorption. This interaction vanishes when the cavity is removed, but also when its front mirror is fully reflecting. The TC model does not distinguish between these situations which, from the perspective of the probe light, correspond to entirely different behaviors. Full-scale numerical simulations of molecular-cavity models provide a viable but costly alternative \cite{32}.

The standard way to analyze experimental results is therefore based on using Fresnel equations together with suitable models for the mirrors and cavity dielectric functions. The molecular subsystem is characterized by Lorentz type function derived from the Bloch equations, where the molecular susceptibility has the form $\chi(\omega)=\ds\frac{A\omega_0^2}{\omega_0^2-\omega^2-i\Gamma\omega}$ \cite{33}. We assume that in this expression, the damping parameter $\Gamma$ determines the width of the molecular resonance occurring at $\omega=\omega_0$ in the absence of the incident light. Thus the resonance width $\Gamma$ by itself does not contain contributions from light dephasing and it cannot   properly describe its effects on transmission and reflection lineshapes. All the more, it cannot address the coherence properties of the transmitted or reflected light as expressed, e.g. by the second order photon correlation function, \cite{34} or the transition from ballistic to diffusive motion of a polariton wavepacket \cite{35}.

 Inside the cavity, the light may suffer  dephasing caused by several physical processes such as thermal fluctuations in the medium filling the cavity interior, acoustic noise, electromagnetic modes already present in the cavity  and some other. In the present paper we generalize the standard Fresnel procedure for calculating transmission and reflection spectra to include the effects of decoherence. For this purpose we invoke a procedure originally developed by Büttiker to account for decoherence in electron transmission through molecular conduction junctions \cite{36,37,39} and widely used to treat incoherent electron \cite{36, 37, 38, 39}, phonon \cite{40, 41, 42} and photon \cite{43} transport as a multichannel scattering problem. 

  We assume that a fraction of light loses its coherence as a result of exchange of photons with a phase randomizing thermal reservoir where temperature is determined by the requirement that the exchange flux vanishes, whereas the remaining light maintains its coherence. Note that this reservoir may not be considered as some specific physical entity. It is introduced merely as means to separate out the dephased fraction of light regardless of the dephasing origin and is an analog to a Büttiker probe used in the theory of electron transport.

\begin{figure}[t] 
\begin{center}
\includegraphics[width=8cm,height=4cm]{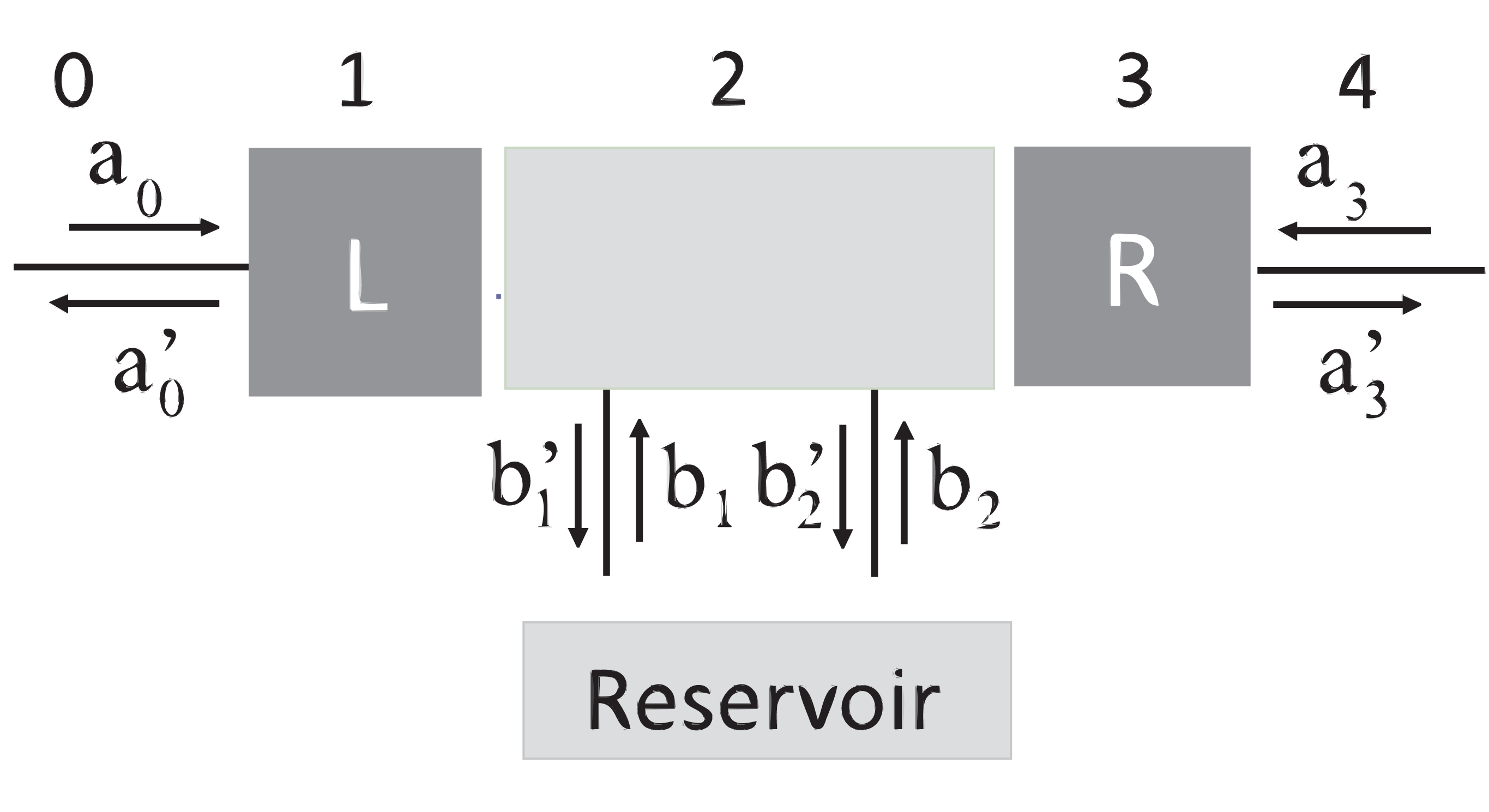} 
\includegraphics[width=5cm,height=3.5cm]{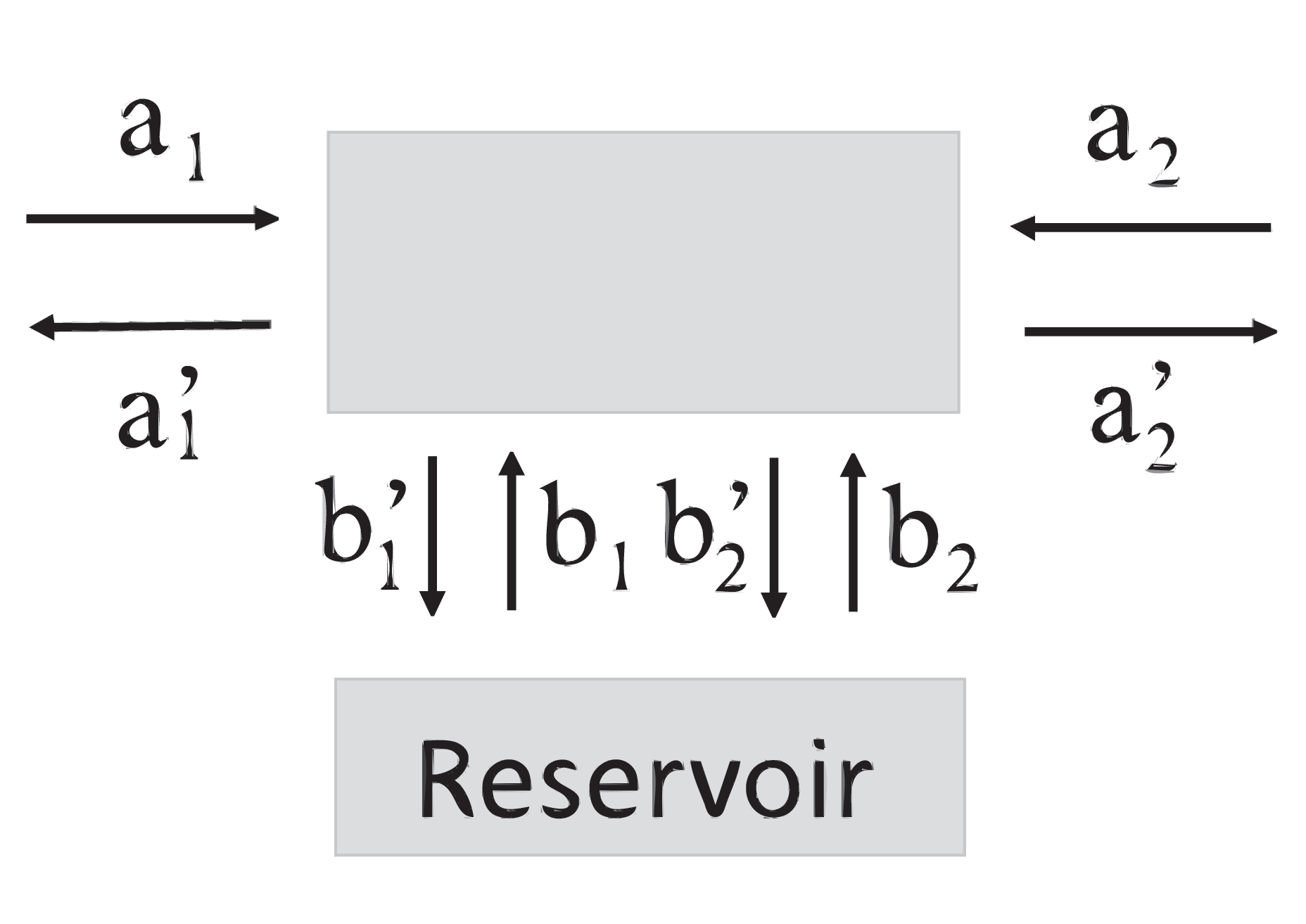}
\caption{Left: Schematic of the considered FP cavity together with the dephasing bath. Dark rectangles represent the mirrors (segments $1$ and $3$) and the light one stands for the space in between (segment $2$). Decoherence is introduced by additional processes, in which light is exchanged with a phase randomizing reservoir via two channels. Right: The cavity interior (that is the space between the mirrors) showing the coupling of the cavity system and the dephasing reservoir.
}
 \label{rateI}
\end{center}
\end{figure}
  The dielectric response of the cavity interior is taken to model a system of identical 2-state molecules and the method is demonstrated for light incident along the normal to the mirrors planes. This leads to modified transmission and reflection coefficients that include the effect of the decoherence induced by this exchange. We find that this decoherence  may strongly affect linear optical response of a FP cavity.
 
\subsection{II. Main equations.}

The considered  model of a FP cavity is presented in Fig.1. Here, the dark rectangles stand for the left and right mirrors, and the light one represents the cavity interior. Assuming that the light propagates along the normal to the mirrors  we may treat it as a combination of the left- and right-traveling waves. The amplitudes of the incident light coming from the left or right are denoted $a_0$  and $a_3$ (typically one of these will be zero) while the corresponding outgoing ( reflection and transmission) amplitudes are labeled $a_0^{'}$ and $a_3^{'}$ (see the Fig.1). Also, in this figure we separately show the cavity interior. The amplitudes coming into and out of the cavity interior through its interfaces with the mirrors are denoted as $a_1$, $a_2$, $a_1^{'}$ and $a_2^{'}$, respectively. To describe possible decoherence effects we invoke a procedure that was applied by Büttiker \cite{36,37,38}. This is done by allowing a fraction of  light in the cavity interior to be exchanged with a phase randomizing reservoir while (a) losing the phase information and (b) keeping the net flux between cavity and reservoir zero. As shown in Fig.1, this exchange is done via two channels  with outgoing amplitudes $b_1$ and $b_2$ leaving the cavity and the incoming amplitudes $b_1{'}$ and $b_2^{'}$ corresponding to dephased  light returning back to the cavity.

 We introduce  two channels for the exchange assuming that each channel maintains the exchange for the light coming into the cavity from a certain (left/right) direction. Together they allow for the complete phase randomization with zero net exchange flux. A similar situation for the case of electron transport was discussed in Ref. \cite{36} and it was shown that a single exchange channel is unable to maintain the total phase randomization for electrons. However, we suppose that in the considered case of  light modes in a FP cavity a single exchange channel may suffice to analyze the decoherence effects, but we do not expect the corresponding modification of the model illustrated in Fig.1 to simplify necessary computations or to bring  some new insights. 

In the absence of dephasing $b_1=b_1^{'}=b_2=b_2^{'}=0$ and the amplitudes $a_0$, $a_0^{'}$, $a_3$, $a_3^{'}$ can be evaluated using Fresnel equations, leading to \cite{44}: 
\begin{gather}
\begin{bmatrix}
a_0\\a_0^{'}
\end{bmatrix}
={\bf M}\begin{bmatrix}
a_3^{'}\\a_3
\end{bmatrix}        \label{1}  
\end{gather}
where:
\be
{\bf M}={\bf D}_0^{-1}{\bf Q}{\bf D}_4 ;   \qquad      {\bf Q}=\Pi_{l=1}^{3}{\bf D}_l{\bf P}_l{\bf D}_{l}^{-1}   \label{2}
\ee
 and the matrices ${\bf D}_l$ ($0\leq l\leq 4$) and ${\bf P}_l$ ($1\leq l\leq 3$) are given by:
\be
{\bf D}_l=\left(\ba{cc}1&1\\n_{l}(\omega)&-n_{l}(\omega);
\\
\ea\right);                     \label{3}
\ee
\be
{\bf P}_l=\left(\ba{cc}\exp(i\Phi_l)&0\\0&\exp(-i\Phi_l)
\\
\ea\right);   \qquad  \Phi_l=\ds\frac{\omega}{c}n_{l}(\omega)L_l.            \label{4}
\ee
In Eqs. (\ref{3}), (\ref{4}) $n_{l}(\omega)=\sqrt{\ds\epsilon_l(\omega)}$ ($0\leq l\leq 4$) are the refraction indices. The subscripts $l=0,4$ represent the outside medium on the left and right of the cavity, $l=1,3$ denote the left and right mirrors and $l=2$ corresponds to the cavity interior. Also, $L_l$ and $L_3$ are the thicknesses of the  left and right mirrors while $L_2$ is the distance between the inner mirrors planes. 


 Eq.(\ref{1}) leads to the standard results for light scattering of a FP cavity (at normal incidence) in the absence of decoherence. Further we take vacuum as the outside media. Thus ${\bf D}_0={\bf D}_4$ and the matrix ${\bf M}$ determined by Eq.(\ref{1}) may be presented as the product  of three $2\times 2$ matrices:
\be
 {\bf M}= {\bf M}^{(1)}{\bf M}^{(2)}{\bf M}^{(3)}             \label{5}
\ee  
 where 
\be
{\bf M}^{(l)}={\bf D}_0^{-1}{\bf D}_l{\bf P}_l{\bf D}_{l}^{-1}{\bf D}_0 ; \qquad (1\leq l\leq 3).        \label{6}
\ee
In Eq.(\ref{5}), the first and the last factors correspond to the mirrors and the middle one is associated with the cavity interior:
\begin{gather}
\begin{bmatrix}
a_0\\a_0^{'}
\end{bmatrix}
={\bf M}^{(1)}\begin{bmatrix}
a_1\\a_1^{'}
\end{bmatrix};      \label{7}    
\end{gather}
\begin{gather}
\begin{bmatrix}
a_1\\a_1^{'}
\end{bmatrix}
={\bf M}^{(2)}\begin{bmatrix}
a_2^{'}\\a_2
\end{bmatrix};     \label{8}       
\end{gather}
\begin{gather}
\begin{bmatrix}
a_2^{'}\\a_2
\end{bmatrix}
={\bf M}^{(3)}\begin{bmatrix}
a_3^{'}\\a_3
\end{bmatrix}.        \label{9}  
\end{gather}
 The matrix elements of the  matrices ${\bf M}^{(l)}$ ($l=1,2,3$) have the form:
\be
M_{11}^{(l)}=\cos(\Phi_{l})+\frac{i}{2}\left(n_{l}+\frac{1}{n_{l}}\right)\sin(\Phi_{l});       \label{10}
\ee
\be
 M_{22}^{(l)}=\cos(\Phi_{l})-\frac{i}{2}\left(n_{l}+\frac{1}{n_{l}}\right)\sin(\Phi_{l});    \label{11}
\ee
\be
M_{12}^{(l)}=-M_{21}^{(l)}=\frac{i}{2}\left(n_{l}-\frac{1}{n_{l}}\right)\sin(\Phi_{l})  \label{12}                                                                                    
\ee
 For identical mirrors ${\bf M}^{(1)}={\bf M}^{(3)}$. Note that Eqs. (\ref{6})-(\ref{9}) express the transmission process in terms of transfer matrices ${\bf M}^{((l)}$. We may express the same information in terms of scattering matrices. For example, Eq.(\ref{8}) may be transformed to the form:
\begin{gather}
\begin{bmatrix}
a_1^{'}\\a_2^{'}
\end{bmatrix}
={\bf s}^{(2)}\begin{bmatrix}
a_1\\a_2
\end{bmatrix}.            \label{13}      
\end{gather}
where the scattering matrix ${\bf s}^{(2)}$ is:
\be
{\bf s}^{(2)}=\left[\ba{cc}r_2&t_2
\\t_2&r_2
\\
\ea\right]              \label{14}
\ee
and $r_2=\ds\frac{M^{(2)}_{21}}{M^{(2)}_{11}}$, $t_2=\ds\frac{1}{M_{11}^{(2)}}$.

 In general, the transfer matrices ${\bf M}^{(l)}$ defined by Eq.(\ref{6}) and the scattering matrix ${\bf s}$ given by Eq.(\ref{8}) are corresponding to unitary operations  only provided that refractive indices $n_l$ are real, and the system does not absorb light. In the considered case all FP cavity subsystems (bounding metallic mirrors and cavity interior filled with molecules) are capable to introduce losses to the light intensity. Correspondingly, all $n_l$ include nonzero imaginary parts and these matrices are not unitary ones. 

To account for decoherence processes undergone by the light inside the cavity we apply a procedure introduced by Büttiker \cite{36}.  Following this procedure, we replace Eq.(\ref{13}) that connects between the incoming and outgoing light amplitudes at the cavity interior-mirrors interfaces) by: 
\begin{gather}
\begin{bmatrix}
a^{'}_1\\a^{'}_2\\b^{'}_1\\b^{'}_2
\end{bmatrix}
={\bf s}\begin{bmatrix}
a_1\\a_2\\b_1\\b_2
\end{bmatrix};                         \label{15}   
\end{gather}
with the $4\times 4$ matrix ${\bf S}$ taken to be:
\be
{\bf S}=\left[\ba{cccc}r_2&\alpha t_2&\beta t_2 &0
\\\alpha t_2&r_2&0&\beta t_2 
\\\beta t_2 &0&r_2&-\alpha t_2
\\0&\beta t_2&-\alpha t_2&r_2
\\
\ea\right].                      \label{16}
\ee
 The parameters $\alpha$ and $\beta$ ($\alpha^2+\beta^2=1$) determine the probabilities for light to avoid or undergo phase destruction process, respectively. Note that similar to Ref.\cite{36} two channels are needed to make the matrix ${\bf S}$ in Eq.(\ref{16}) unitary in the limit when the system's refractive indices are real and light energy is conserved.

 Next, Eqs.(\ref{7})-(\ref{9}), (\ref{15}) and (\ref{16}) are used to find the direct relationship between the amplitudes $a_0$, $a_3$, $b_1$, $b_2$ incoming from the left and right environment and from the decoherence bath and the amplitudes  $a^{'}_0$, $a^{'}_3$, $b^{'}_1$ and $b^{'}_2$ outgoing into these environments. The details of the calculations are given in Appendix A. This calculation leads to:
\begin{gather}
\begin{bmatrix}
a^{'}_0\\a^{'}_3\\b^{'}_1\\b^{'}_2
\end{bmatrix}
={\bf T}\begin{bmatrix}
a_0\\a_3\\b_1\\b_2
\end{bmatrix}.              \label{17}   
\end{gather}
 where the matrix ${\bf T}$ has the form:
\be
{\bf T}=\left[\ba{cccc}\tilde{r}&\tilde{t}&\beta \vartheta &-\alpha\beta\tau
\\\tilde{t}&\tilde{r}&-\alpha\beta\tau&\beta \vartheta 
\\\beta\vartheta &-\alpha\beta\tau&\beta^2\eta&-\alpha\Theta
\\-\alpha\beta\tau&\beta\vartheta &-\alpha\Theta&\beta^2\eta
\\
\ea\right].                      \label{18}
\ee
  Expressions for the terms $\tilde{r}$, $\tilde{t}$, $\vartheta$, $\eta$, $\rho$, $\tau$ and $\Theta$ are given by Eqs. (\ref{A11}), (\ref{A12}). Note that $\tilde{t}$ is proportional to $\alpha$, thus this term becomes zero in the limit of strong phase randomization.

In the limit $\beta=0$, ${\bf T}$ becomes a block matrix: ${\bf T}={\bf T}_1\oplus {\bf T}_2$ where only first block that relates $a^{'}_0$, $a^{'}_3$ to $a_0$, $a_3$ is meaningful. We get:
\begin{gather}
\begin{bmatrix}
a^{'}_0\\a^{'}_3
\end{bmatrix}
=\left[\ba{cc}\tilde{r}&\tilde{t}
\\\tilde{t}&\tilde{r}
\\
\ea\right]
\begin{bmatrix}
a_0\\a_3
\end{bmatrix}        \label{19}  
\end{gather}
which is equivalent to Eq.(\ref{1}). Assuming that the light comes from the left ($a_3=0$) this leads to $\tilde{r}=\ds\frac{M_{21}}{M_{11}}$, $\tilde{t}=\ds\frac{1}{M_{11}}$ which determine the reflection and transmission in the coherent limit. 

In general, the matrix elements $T_{ik}$ provide probabilities for various scenarios involving the behavior of light in the cavity. For instance, the light  amplitude $a_0$ coming from the left may be reflected back with the probability $|\tilde{r}|^2$,  coherently transmitted through the system with the probability $|\tilde{t|}^2$ or scattered into the dephasing reservoir via the channels $1$ or $2$ with the probabilities $\beta^2 |\vartheta|^2$ or $\alpha^2\beta^2|\tau|^2$. In the considered steady state regime light intensities are proportional to the corresponding squared amplitudes and obey the equations similar to that derived for the case of electron transport through molecules \cite{38,39,45}. Using these equations and assuming that the net exchange flux with the decoherence reservoir vanishes \cite{46}, we may derive  (Appendix B) the expressions for the optical transmission and reflection influenced by the effects of decoherence:
\be
T=\bigg|\frac{a_3^{'}}{a_0}\bigg|^2=W_{21}+\frac{1}{2}\frac{(W_{13}+W_{14})^2}{1-W_{33}-W_{44}}                     \label{20}
\ee
\be
R=\bigg|\frac{a_0^{'}}{a_0}\bigg|^2=W_{11}+\frac{1}{2}\frac{(W_{13}+W_{14})^2}{1-W_{33}-W_{44}}                     \label{21}
\ee
where $W_{ik}=|T_{ik}|^2$.  
 When $T$ and $R$ are known, the light extinction $A$ may be also found: $A=1-R-T$.  In the absence of decoherence ($\alpha=1$) the extinction  solely results from absorption. Then the transmission $T$ and the reflection $R$ are respectively equal to $|\tilde {t}|^2=\ds\frac{1}{|M_{11}|^2}$ and $|\tilde{r}|^2=\ds\frac{|M_{21}|^2}{|M_{11}|^2}$ which are the well known results that follow from Eq.(\ref{1}). 

\subsection{III. Results and discussion}

In the following analysis we take the mirrors to be identical $(L_1=L_3=d)$, $n_3(\omega)=n_1(\omega)=\sqrt{\epsilon_{1}(\omega)}$, $\Phi_{1}=\Phi_3=\ds\frac{\omega}{c}n_{1}(\omega)d$. We also take vacuum as the outside media so that $n_0=n_4=1$. Hence ${\bf D}_1={\bf D}_3$  and ${\bf D}_0={\bf D}_4$. For the dielectric function of the mirrors we use the Drude form $\ds\left(\epsilon_{1}(\omega)=\epsilon_{\infty}-\ds\frac{\omega_p^2}{\omega(\omega+i\gamma)}\right)$, with the plasma frequency of the metal $\omega_p$ and the high frequency contribution from atomic core electrons $\epsilon_{\infty}$. Finally, for the molecular environment we take $\epsilon_2(\omega)=1+\chi(\omega)$ where $\chi(\omega)=\ds\frac{A\omega_0^2}{\omega_0^2-\omega^2-i\Gamma\omega}$. Throughout the calculations we use:  $\omega_p=9$eV, $\epsilon_{\infty}=9.2$ (which are close to the corresponding characteristics for gold) and $\gamma=0.2$eV. Also, we assume that $\Gamma=0.08$eV and $\omega_0=1.09$eV characterizing the unspecified two-level molecules placed in the cavity coincides with the frequency of the fundamental cavity mode at the chosen distance between the mirrors ($L_2=500$nm). The dimensionless factor $A$ depends linearly on on the number $N$ of molecules involved in light-matter interaction. In the present calculations we put $A=2\times 10^{-3}$. Assuming that the coupling strength between a single molecule and the light $g\sim 10^{-7}$eV we may estimate $N\sim 10^{11}$ to obtain  for Rabi frequency a value close to that used in the present work.
\begin{figure}[t] 
\begin{center}
\includegraphics[width=4cm,height=3.5cm]{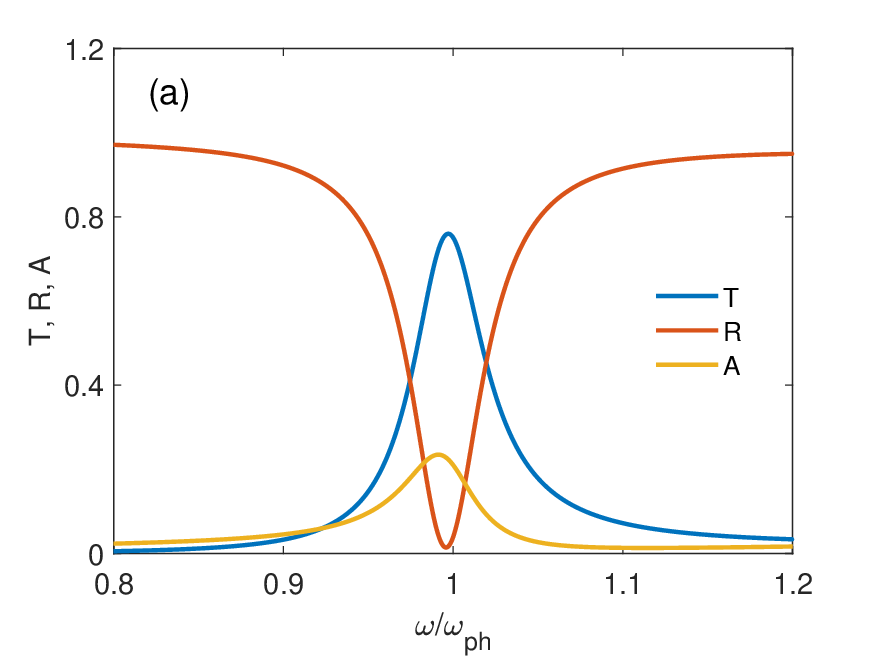} 
\includegraphics[width=4cm,height=3.5cm]{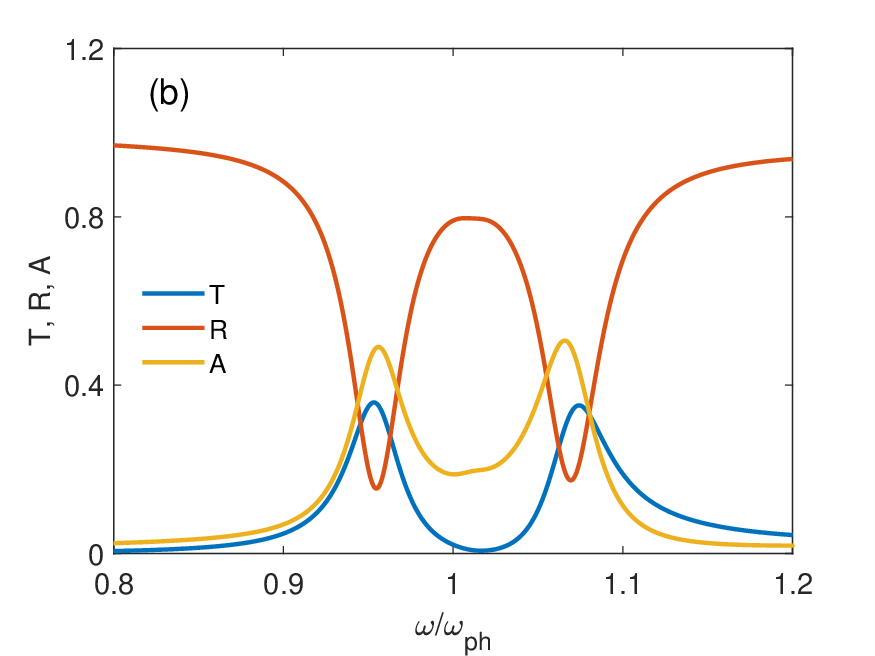}
\caption{ Transmission ($T$), reflection ($R)$ and extinction $(A)$ as functions of the light frequency $\omega$ in the coherent limit ($\beta=0$)in the absence (a) and in the presence (b) of molecular medium in the cavity interior. The curves are plotted at $\omega_p=9$eV, $\epsilon_{\infty}=9.2$, $\gamma=0.2$eV, $d=60$nm, $L_2=500$nm and the cavity mode frequency $\omega_{ph}=1.09$eV. The refraction index $n_2(\omega)=1$ (a) and $n_2(\omega)=\sqrt{1+\ds\frac{0.5\Omega^2}{\omega_0^2-\omega^2-i\Gamma\omega}}$ (b) where $\omega_0=1.09$eV, $\Gamma=0.08$eV and Rabi frequency $\Omega=0.1$eV. 
}
 \label{rateI}
\end{center}\end{figure}

Fig.2 shows the standard spectra obtained in the absence of decoherence when the light extinction solely originates from its absorption. The panel (a) displays the light transmission, reflection and absorption in a FP cavity free from molecules while the panel (b) demonstrates the effect of molecules on these spectra. A single peak in transmission  at $\omega=\omega_{ph}$ associated with the cavity mode of the empty cavity shown in Fig.2(a) is replaced by two polariton peaks (separated by the Rabi splitting) resulting from the interaction between the molecular resonance and this mode displayed in Fig.2(b). The polariton peaks appear as a result of strong light-matter coupling at the frequencies $\omega_{\pm}=\ds\frac{1}{2}\left(\omega_{ph}+\omega_0\pm\sqrt{(\omega_{ph}-\omega_0)^2+\ds\frac{1}{4}\Omega^2}\right)$, $\Omega$ being the Rabi frequency. 
\begin{figure}[t] 
\begin{center}
\includegraphics[width=4cm,height=3.5cm]{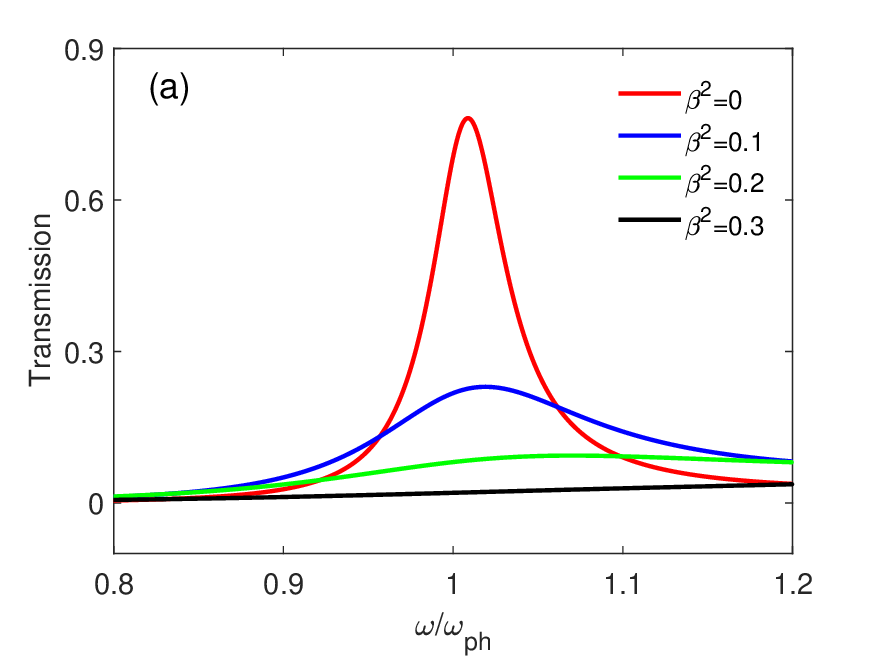} 
\includegraphics[width=4cm,height=3.5cm]{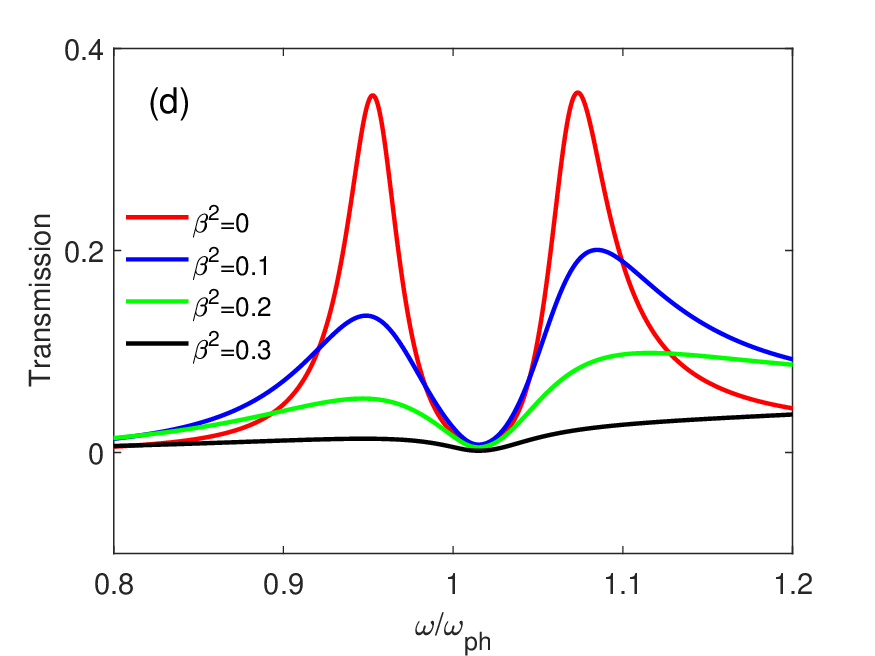}
\includegraphics[width=4cm,height=3.5cm]{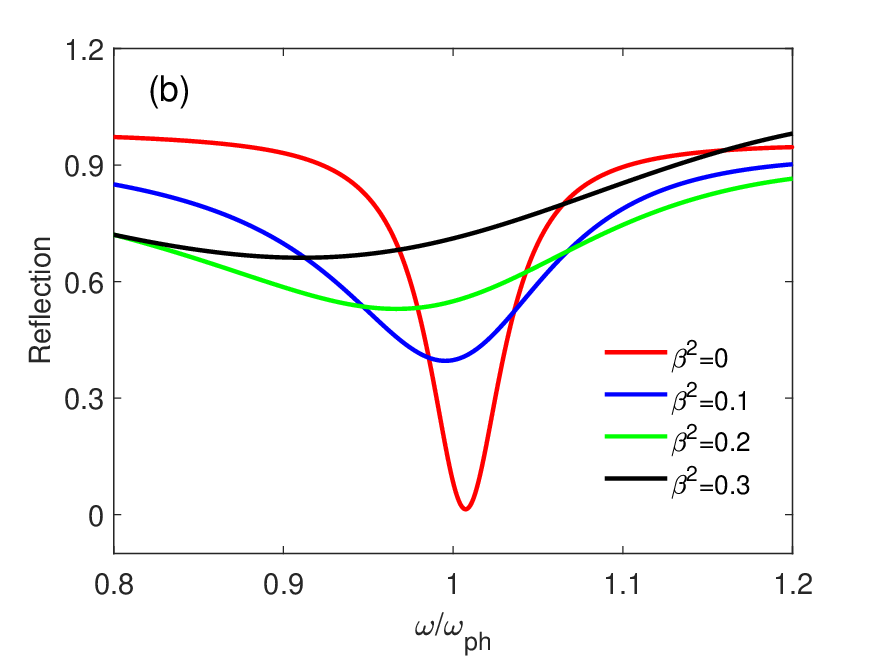}
\includegraphics[width=4cm,height=3.5cm]{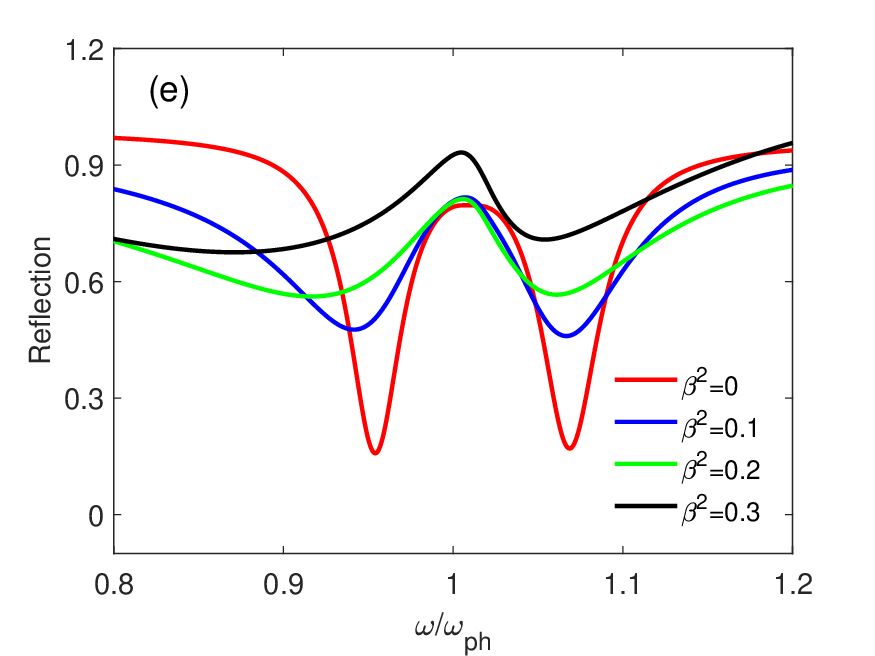}
\includegraphics[width=4cm,height=3.5cm]{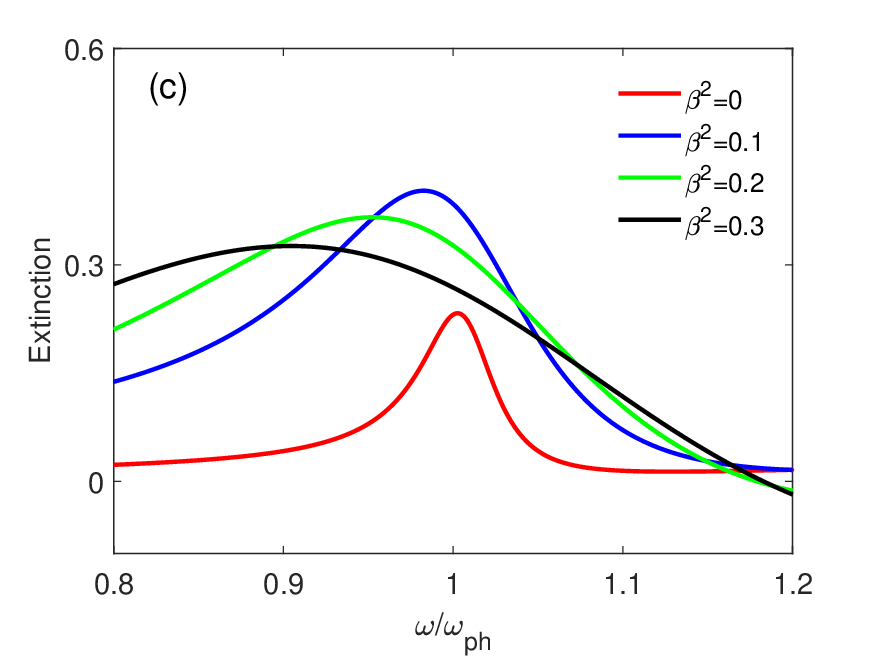}
\includegraphics[width=4cm,height=3.5cm]{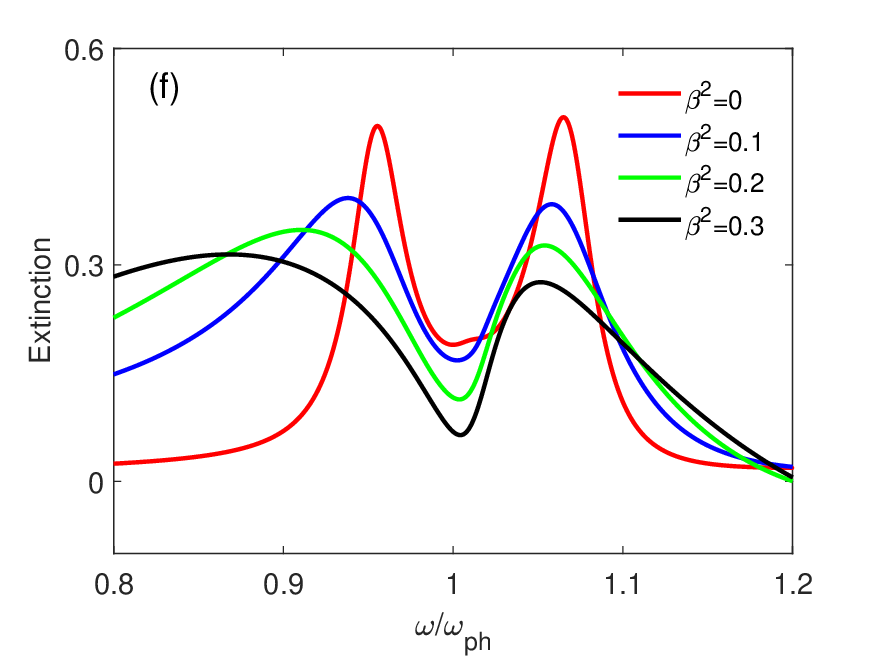}
\caption{ The effect of decoherence on the light transmission, reflection and extinction spectra of a FP cavity in the absence (panels (a), (b), (c))) and in the presence (panels (d), (e), (f)) of molecules in the cavity interior. All parameters except $\beta$ used here are taken the same as in Fig.2.
}
 \label{rateI}
\end{center}\end{figure} 

To study the the effects of decoherence we compute the optical characteristics at several values of the parameter $\beta$. Obtained results are displayed in Fig.3. They show that even moderate decoherence strongly affects the cavity optical spectra. In both the empty cavity (panel (a)) and the cavity whose interior is filled with molecules (panel (d)), the transmission dramatically drops as the  decoherence effects wash out of the cavity mode. The polariton peaks in the transmission are nearly erased at $\beta^2=0.2$. In contrast, the decoherence leads to the growth of the reflection (panels (b) and (e)). Nevertheless, the dips in the reflection spectra become more shallow as the dephasing intensifies and the signatures of molecular polaritons gradually vanish. The dephasing reduces the extinction of the incoming light near the polariton frequencies $\omega_{\pm}$ and promotes it away from the latter, as shown in the panel (f). Also, the decoherence enhances the light extinction in the empty cavity (see panel (c)).  As the decoherence strengthens, the extinction peaks displayed in this panel broaden and show the increasing red shift. One observes a similar red shift of the reflection dips shown in the panel (b). We conjecture that this happens because of the combined effect of the fading away cavity mode and the properties of the bounding mirrors. At low frequencies ($\omega\ll\omega_{ph})$ reflection is mostly determined by the golden bounding mirrors and decreases as the light frequency rises. When $\omega$ approaches $\omega_{ph}$ the effect of the cavity mode becomes important and causes the minimum in the reflection. The broadening of the reflection dip associated with the cavity mode shifts this minimum to lower frequencies. Also, at low frequencies we may neglect the light transmission and assume that $A=1-R$. Thus, in the presence of decoherence, the light extinction may be expected to exhibit a broad maximum at low light frequencies showing the red shift as the decoherence strengthens. Note that this feature may disappear if gold is replaced by another material (e.g. silver) which shows a different frequency dependence of light reflection.	
			
In general, light extinction results from the combined effect of light absorption and scattering. At negligible decoherence the extinction is reduced to the light absorption by the cavity mode and exhibits a peak at $\omega=\omega_{ph}$. However, even in the presence of a moderate decoherence the contribution to the extinction from scattering becomes very significant and  away from $\omega=\omega_{ph}$ it predominates.
To  further elucidate the contribution to the extinction from the  light scattering we consider the case when the mirrors refractive indices $n_{1,3}$ are real  ($\gamma=0$), so that the mirrors do not absorb light. In this limit in the absence of molecules the extinction is determined by  light scattering and becomes zero in the coherent limit, as seen in Fig.4(a). Comparing full and dashed lines in Fig.4(a) we see that light absorption by the mirrors makes only a small contribution to the extinction, except when $\beta=0$. The same is true in the presence of molecules, as seen in Fig.4(b). 

Figs.3 and 4 show that the effect of decoherence on the linear cavity spectra is mainly to broaden the spectral features associated with the cavity mode and (in the presence of molecules) the polariton features. This broadening adds to other broadening effects originated from finite lifetimes of the cavity mode (controlled by the mirrors thickness and their dielectric properties) and from the molecular excitations. Further broadening is caused by dynamic and static disorder in the molecular subsystem \cite{47,48}. Fig.5 shows that broadening caused by  light phase breaking, as introduced here, does not depend on the Rabi splitting associated with the molecules coupling with the cavity mode, which is a characteristic behavior of homogeneous broadening (fast disorder limit) \cite{48}.
\begin{figure}[t] 
\begin{center}
\includegraphics[width=4cm,height=3.5cm]{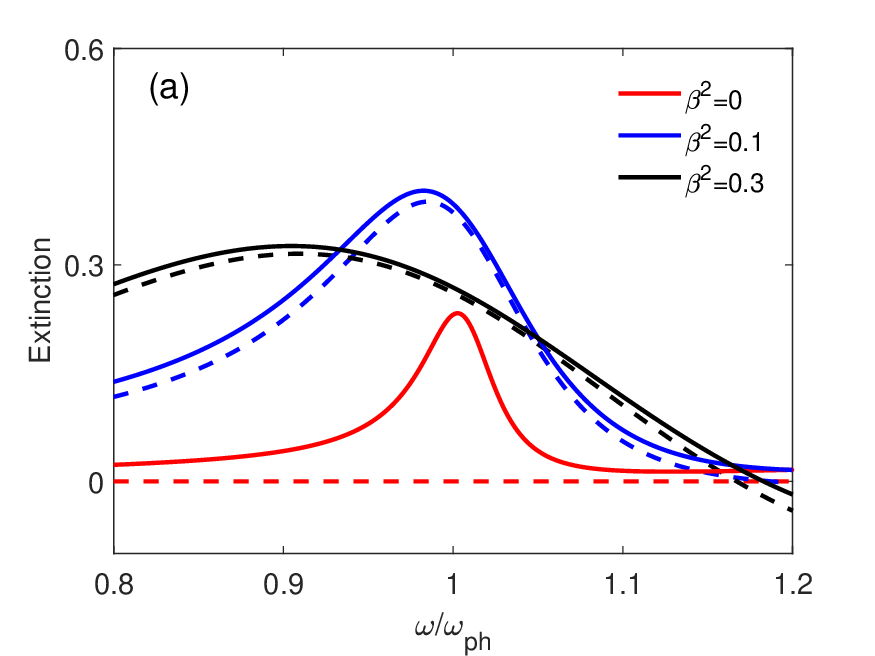} 
\includegraphics[width=4cm,height=3.5cm]{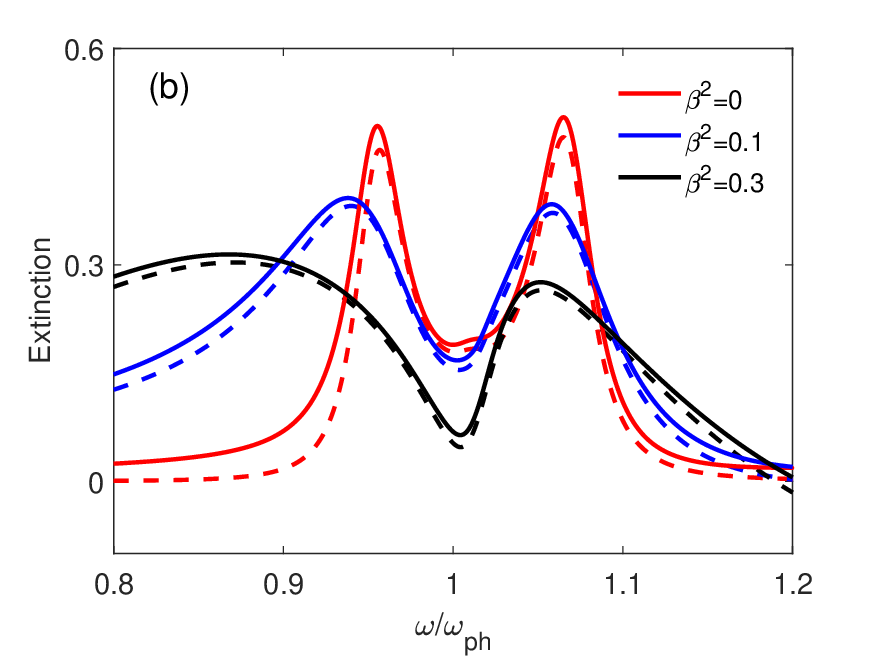}
\caption{ The effect of decoherence on the light extinction in the empty FP cavity (a) and the cavity filled with molecules (b). Dashed curves are plotted neglecting the  absorption in the bounding mirrors and solid ones are plotted assuming that the mirrors absorb the light.
}
 \label{rateI}
\end{center}\end{figure} 
\begin{figure}[t] 
\begin{center}
\includegraphics[width=6cm,height=4.5cm]{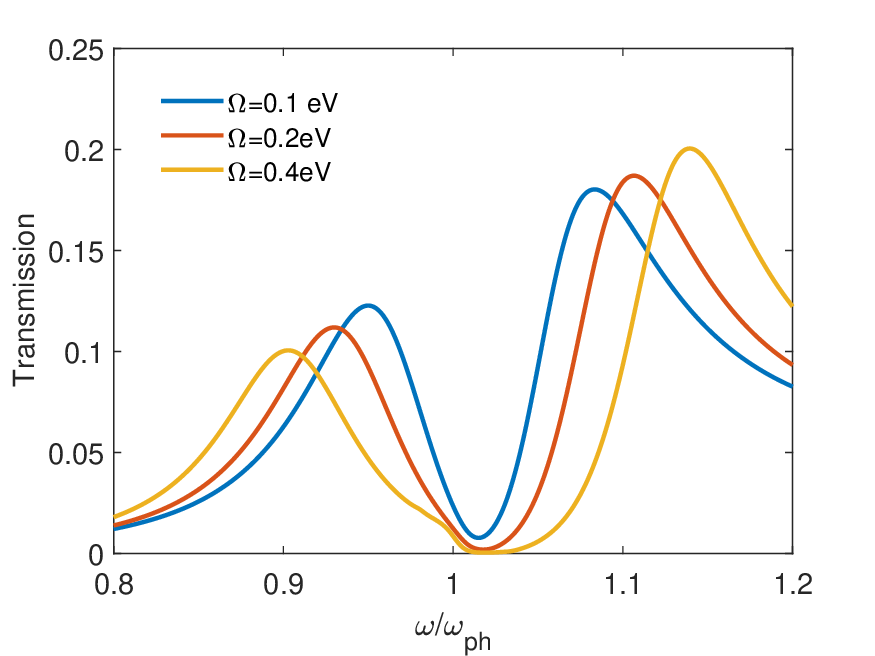} 
\caption{ The effect of light-matter interaction strength on the light transmission in the presence of decoherence. The displayed curves are plotted at $\beta^2=0.1$ 
}
 \label{rateI}
\end{center}\end{figure}

\subsection {Conclusions}

In the present work we have treated certain aspects of the molecular polariton problem using an heuristic approach based on the scattering matrix. The research was concentrated on the exploration of possible effects of the  light phase breaking which may occur as a result of the light scattering in the interior of a FP cavity. We studied the effects of decoherence on the optical spectra of FP cavities when the space between the inner surfaces of the bounding mirrors is empty and when it is filled with a molecular medium characterized by a Lorenz-type dielectric response that does not take into account molecular vibrations. (The effect of such vibrations may be taken into account by choosing an appropriate expression for $\chi(\omega)$.)

 In our analysis we employed Fresnel equations combined with the semiclassical Büttiker formalism adapted for  light transport. It is known that the effects of  light decoherence in linear optics were analyzed basing on classical electrodynamics before. For example, a well-defined procedure to treat coherent and incoherent light propagation in layered materials was suggested in Ref.\cite{49}. However, in that work, as well as is some others, the decoherence effects were introduced by means of varying the refractive indices associated with different layers. The novelty and advantage of the present approach is that here we treat the FP cavity interiror as a single layer characterized by a single refractive index but the light splits in fractions in such a way that one of these fractions suffers the phase randomization and another one retains coherence. Varying the relative intensities of these fractions the extent of decoherence effects in the optical spectra of FP cavities may be easily varied and studied using much simpler computational procedure. Note that the present results do not contradict those reported in earlier works. Although we do not include a decoherence contribution into the linewidth of molecular resonances $\Gamma$, all features in the spectra shown in Fig.3 demonstrate the broadening accompanying the phase randomization of light. Thus the present approach allows to catch this effect as well as previously employed methods. Also, we may expect that the obtained results will qualitatively agree with those obtained using quantum input-output theory \cite{50} in the case when the molecules in the cavity are treated as simple two level systems.

It was shown that the optical spectra of the FP cavities are very sensitive to the effects of decoherence. Even moderate light phase breaking affects the cavity mode and results in gradual washing out of the polaritons peaks in the absorption and transmission and  matching dips in the reflection. The transmission spectra are the most affected. When the probability for dephasing of light in the FP cavity interior is still rather small ($\beta^2\approx 0.2$) the transmission becomes close to zero and the molecular polaritons signatures completely disappear, while both reflection (and extinction away from $\omega=\omega_{ph}$) increase. 
  Decoherence processes were seen to significantly change both the heights and widths of polariton peaks. At the same time,the decoherence-induced lines broadening was seen not to depend on the Rabi splitting, as is typical for homogeneous broadening effects. 
	
	While the present work has focused on the effect of decoherence of optical cavities, the corresponding effects on excitation transport is also of interest. The most prominent manifestation of decoherence in junction electronic transport is the transition from ballistic to diffusive motion. \cite{35,51}. The aplication of the present formalism to polariton transport, as observed e.g. in Ref.\cite{34} will be a subject for future study. Note, that implementation of the suggested approach in a 1-dimensional geometry (e.g. a polariton moving with a uniform wavefront parallel to planar cavity mirrors or along a linear waveguide) is similar to that done in many works on electron transport along 1-dimentional molecular chains. Dephasing could be provided by a chain of local independent Buttiker probes (each analogous to the reservoir of Fig.1) distributed along the 1-dimensional path, and effectively transforming coherent Schrodinger propagation to diffusive transport (see e.g \cite{37,38,39,52,53}). Our work in this direction will start by demonstrating that the same can be achieved with Maxwell propagation. The real challenge is to extend this approach to higher dimensions.  In principle, the idea of affecting dephasing using a distribution of Buttiker probes (each affecting local decoherence) can be done in any number of dimensions, as shown in Refs.\cite{50,54} but the numerical effort (which depends on the grid used for such local dephasing agents) and the performance of such an approach will require careful assessment.

\subsection{Declaration of competing interest}

Authors declare that they have no competing financial interests or personal relationships which could influence the work reported in this paper.

\subsection{Data availability statement}

Data sharing is not applicable as no data are created in this study.

\subsection{Acknowledgments}

The present work was supported by the U.S National Science Foundation (DMR-PREM 2122102) and by the European Research Council (ERC) under the European Union's Synergy Program (grant agreement No. 10095861). NZ thanks to Dr. G.M. Zimbovskiy for help in the manuscripts preparation.

\begin{widetext}
\begin {appendix}
\subsection{Appendix A}
\setcounter{equation}{0}
\renewcommand\theequation{A.\arabic{equation}} 

Separating out the interior part of the FP cavity (see Fig.1) and assuming that the  light coming to the dephasing reservoir via one of the channels  cannot reappear in the cavity interior via another channel we can present the relationship between the outgoing amplitudes $a_1^{'}$, $a_2^{'}$, $b_1^{'}$, $b_2^{'}$ and the incoming ones $a_1$, $a_2$, $b_1$, $b_2$ in the form:
\begin{gather}
\begin{bmatrix}
a^{'}_1\\a^{'}_2\\b^{'}_1\\b^{'}_2
\end{bmatrix}.
={\bf s}\begin{bmatrix}
a_1\\a_2\\b_1\\b_2
\end{bmatrix};                         \label{A1}   
\end{gather}
where
\be
{\bf s}=\left[\ba{cccc}r_2&\alpha t_2&\beta t_2&0
\\\alpha t_2&r_2&0&\beta t_2 
\\\beta t_2&0&r_2&-\alpha t_2
\\0&\beta t_2&-\alpha t_2&r_2
\\
\ea\right].                      \label{A2}
\ee
 Using the equations:
\begin{gather}
\begin{bmatrix}
a_0\\a_0^{'}
\end{bmatrix}
={\bf M}^{(1)}\begin{bmatrix}
a_1\\a_1^{'}
\end{bmatrix}          
\qquad
\begin{bmatrix}
a_2^{'}\\a_2
\end{bmatrix}
={\bf M}^{(1)}\begin{bmatrix}
a_3^{'}\\a_3
\end{bmatrix}        \label{A3}  
\end{gather}
we get $\left(\nu=\ds\frac{M_{22}^{(1)}}{M_{11}^{(1)}}\right)$:
\be
a_1=\frac{1}{t_1}\left(\nu a_0+r_1a^{'}_0\right);    \qquad    a^{'}_1=\frac{1}{t_1}\left(-r_1 a_0+ a^{'}_0\right);            \label{A4}
\ee
and
\be
a_2=\frac{1}{t_1}\left(\nu a_3+r_1 a^{'}_3\right);    \qquad   a_2^{'}=\frac{1}{t_1}\left(-r_1a_3+ a^{'}_3\right).               \label{A5}
\ee
Also, using Eqs.(\ref{A1})-(\ref{A2}) we may write:
\be
a^{'}_1=r_2a_1+\alpha t_2a_2+\beta b_1;   \qquad   a^{'}_2=\alpha t_2a_1+r_2a_2+\beta b_2.                                     \label{A6}
\ee            
Combining Eqs. (\ref{A4})-(\ref{A6}) we arrive at the expressions:
\be
a^{'}_0=\frac{1}{\Delta}\left(\Lambda a_0+\alpha t_1^2 t_2 a_3+\beta t_1t_2(1-r_1r_2)b_1-\alpha\beta t_1t_2^2 r_1 b_2\right);                 \label{A7}
\ee
\be
a^{'}_3=\frac{1}{\Delta}\left(\alpha t_1^2t_2a_0+\Lambda a_3-\alpha\beta t_1t_2^2r_1b_1+\beta t_1t_2(1-r_1r_2)b_2\right);                    \label{A8}
\ee
\be
b^{'}_1=\frac{1}{\Delta}\left(\beta t_1t_2(1-r_1r_2)a_0-\alpha\beta t_1t_2^2 r_1 a_3+\beta^2\rho b_1-\alpha t_2\theta b_2\right); \label{A9}
\ee
\be
b^{'}_2=\frac{1}{\Delta}\left(-\alpha\beta t_1t_2^2r_1 a_0+\beta t_1t_2(1-r_1r_2)a_3-\alpha t_2\theta b_1+\rho b_2\right);  \label{A10}
\ee
where
$\Delta=(1-r_1r_2)^2-\alpha^2 t_2^2 r_1^2$, $\Lambda=(1-r_1r_2)(r_1+r_2\nu)+\alpha^2t_2^2r_1\nu$, $\rho=t_2^2(1-r_1r_2)+r_2\Delta)$, $\theta=\left(\beta^2 t_2^2r_1^2+\Delta\right)$.  Introducing extra denotations, namely:
\be
\tilde{r}=\frac{\Lambda}{\Delta};   \qquad  \tilde{t}=\frac{\alpha t_1^2t_2}{\Delta}; \qquad  \tau=\frac{t_1 t_2^2 r_1}{\Delta};  \label{A11}
\ee
\be
\vartheta=\frac{t_1t_2(1-r_1r_2)}{\Delta}; \qquad    \Theta=\frac{t_2\theta}{\Delta};  \qquad \eta=\frac{\rho}{\Delta}  \label{A12}
\ee
  we get the matrix ${\bf T}$ in the form:
\be
{\bf T}=\left[\ba{cccc}\tilde{r}&\tilde{t}&\beta \vartheta &-\alpha\beta\tau
\\\tilde{t}&\tilde{r}&-\alpha\beta\tau&\beta \vartheta 
\\\beta\vartheta &-\alpha\beta\tau&\beta^2\eta&-\alpha\Theta
\\-\alpha\beta\tau&\beta\vartheta &-\alpha\Theta&\beta^2\eta
\\
\ea\right].                      \label{A13}
\ee
 The quantities $\tilde{t}$ and $\tilde{r}$ determine the transmission and reflection through the cavity. In the coherent limit ($\alpha=1$, $\beta=0$), $\tilde{t}$ and $\tilde{r}$ are given by Eq.(\ref{1}), namely, $\tilde{t}=\ds\frac{1}{M_{11}}$ and $\tilde{r}=\ds\frac{M_{21}}{M_{11}}$. 
\end{appendix}
\begin {appendix}
\subsection{Appendix B}
\setcounter{equation}{0}
\renewcommand\theequation{B.\arabic{equation}}
 
 We introduce outgoing from the cavity fluxes $J_i^{'}$ and incoming fluxes $J_k$ $(1\leq i,k\leq 4)$ which are proportional to the corresponding squared amplitudes, namely: $J_1^{'}\propto |a_0^{'}|^2$, $J_{2}^{'}\propto |a_{3}^{'}|^2$, $J_{3,4}^{'}\propto |b_{1,2}^{'}|^2$, $J_1\propto |a_0|^2$ and so on. Following the way used to explore electron transport through molecules \cite{36, 38,45} we write the relationships between the outgoing and the incoming fluxes:
\be
J_i^{'}=\sum_k W_{ik}J_k ;            \label{B1}
\ee        
where $W_{ik}=|T_{ik}|^2$.
To provide zero exchange flux  we require that $J^{'}_3+J^{'}_4=J_3+J_4$. Then, assuming that $a_3=0$ and summing up Eqs.(\ref{B1}) and using the symmetry of the matrix ${\bf W}$ ($W_{ik}=W_{ki})$  we get:
\be
J_3+J_4=J_1\frac{W_{31}+W_{41}}{2-W_{33}-W_{44}-2W_{34}}             \label{B2}
\ee
Substituting these results into the equations:
\be
J^{'}_1=W_{11}J_1+W_{13}J_3+W_{14}J_4;    \qquad      J_4^{'}=W_{21}J_1+W_{23}J_3+W_{24}J_4           \label{B3}
\ee
Assuming that $J_3=J_4$ and using the relationships which follow from the form of the scattering matrix ${\bf T}$ ($T_{13}=T_{24}$, $T_{14}=T_{23}$) 
we get the following expressions for the reflection and transmission:
\be
R=\bigg|\frac{a_0^{'}}{a_0}\bigg|=\frac{J_1^{'}}{J_1}=W_{11}+\frac{1}{2}\frac{(W_{13}+W_{14})^2}{1-W_{33}-W_{34}}      \label{B4}
\ee
\be
T=\bigg|\frac{a_3^{'}}{a_0}\bigg|=\frac{J_4^{'}}{J_1}=W_{21}+\frac{1}{2}\frac{(W_{13}+W_{14})^2}{1-W_{33}-W_{34}} 
\ee                   
\end{appendix}

\end{widetext}



\begin{thebibliography}{99}
\bibitem{46} Note that this requirement, namely $J_3+J_4-J_3^{'}-J_4^{'} = 0, $ where the relevant fluxes are defined in Appendix B can be used to determine the temperature of Bose-Einstein distribution characterizing the dephasing reservoir.
\end{thebibliography}
\end{document}